\documentclass[twocolumn,showpacs,aps,prl,superscriptaddress]{revtex4}

\usepackage{xspace}
\usepackage{graphicx}
\usepackage{dcolumn}
\usepackage{amsmath}
\usepackage{epsfig}
\usepackage{multirow}

\input babarsym.tex

\newcommand{\ntaupair}         {\ensuremath{3.53\times 10^8}\xspace}
\newcommand{\lumi}             {\ensuremath{384.1\invfb}\xspace}
\newcommand{\lumion}           {\ensuremath{347.5 \invfb}\xspace}
\newcommand{\lumioff}          {\ensuremath{36.6 \invfb}\xspace}

\newcommand{\Mnu}              {\ensuremath{{m_\nu}^2}\xspace}

\newcommand{\ptmiss}           {\ensuremath{p^{T}_{\mathrm{miss}}}\xspace}
\newcommand{\logmisspt}        {\ensuremath{p^T_{\mathrm{miss}}/\sqrt{s}}\xspace}

\newcommand{\tagegam}          {\ensuremath{\Sigma E^{CM}_{\mathrm{neutral}}}\xspace}

\def\DeltaE                    {\ensuremath{\Delta E}\xspace}

\newcommand{\eett}             {\ensuremath{e^+e^- \to \tautau}\xspace}

\def\eff                       {\ensuremath{\varepsilon}\xspace}

\def\tenseven                  {\ensuremath{\times 10^{-7}}\xspace}

\def\tautolome                 {\ensuremath{\tau^\pm \to \ell^\pm\omega}\xspace}
\def\tautomome                  {\ensuremath{\tau^\pm \to \mu^\pm \omega}\xspace}
\def\tautoeome                 {\ensuremath{\tau^\pm \to \electron^\pm \omega}\xspace}
\def\BRtaueome                 {\ensuremath{\BR(\tau^\pm \to \electron^\pm \omega)}\xspace}
\def\BRtaumome                 {\ensuremath{\BR(\tau^\pm \to \mu^\pm \omega)}\xspace}
\def\ptogg   {\ensuremath{\piz \to \gaga}\xspace}
\def\omeppp  {\ensuremath{\omega \to \pi^+\pi^-\piz}\xspace}
\def\ppp  {\ensuremath{\pi^+\pi^-\piz}}
\newcommand{\UpperLimiteome}   {\BRtaueome$<1.1\times10^{-7}$\xspace}
\newcommand{\UpperLimitmome}   {\BRtaumome$<1.0\times10^{-7}$\xspace}

\newcommand{\tautoomega}  {\ensuremath{\tau^- \to 2\pi^-\pi^+\piz\nu}\xspace}

\newcommand{\roots}        {\ensuremath{\sqrt{s}}\xspace}

\def\kk         {\mbox{\tt KK2F}\xspace}
\def\tauola     {\mbox{\tt TAUOLA}\xspace}
\def\photos     {\mbox{\tt PHOTOS}\xspace}

\def\evtgen     {\mbox{\tt EVTGEN}\xspace}
\def\jetset     {\mbox{\tt JETSET}\xspace}
\def\geant      {\mbox{\tt GEANT4}\xspace}

\newcommand{\gevccgevcc}{\ensuremath{{\mathrm{\,Ge\kern -0.1em V^2\!/}c^4}}\xspace}
\newcommand{\evcc}{\ensuremath{{\mathrm{\,e\kern -0.1em V\!/}c^2}}\xspace}
\newcommand{\CM} {\mbox{c.m.}\xspace}

\newcommand{\BABARPubYear}     {07}
\newcommand{\BABARPubNumber}  {064}

\newcommand{\SLACPubNumber} {12971}

\def\figurebox#1#2#3{%
    \def\arg{#3}%
    \ifx\arg\empty
    {\hfill\vbox{\hsize#2\hrule\hbox to #2{\vrule\hfill\vbox to #1{\hsize#2\vfill}\vrule}\hrule}\hfill}%
    \else
    {\hfill\epsfbox{#3}\hfill}%
    \fi}

\begin{document}

\preprint{\babar-PUB-\BABARPubYear/\BABARPubNumber} 
\preprint{SLAC-PUB-\SLACPubNumber} 

\begin{flushleft}
\begin{tabular}{lcr}
\babar-PUB-\BABARPubYear/\BABARPubNumber \\
SLAC-PUB-\SLACPubNumber\\
\end{tabular}\\
\end{flushleft}

\title{
{\large \bf \boldmath Search for Lepton Flavor Violating Decays \tautolome~($\ell = e, \mu$) }
}

%
\author{B.~Aubert}
\author{M.~Bona}
\author{Y.~Karyotakis}
\author{J.~P.~Lees}
\author{V.~Poireau}
\author{X.~Prudent}
\author{V.~Tisserand}
\author{A.~Zghiche}
\affiliation{Laboratoire de Physique des Particules, IN2P3/CNRS et Universit\'e de Savoie, F-74941 Annecy-Le-Vieux, France }
\author{J.~Garra~Tico}
\author{E.~Grauges}
\affiliation{Universitat de Barcelona, Facultat de Fisica, Departament ECM, E-08028 Barcelona, Spain }
% ...
\collaboration{The \babar\ Collaboration}
\noaffiliation

\date{\today}
\begin{abstract}
A search for lepton flavor violating decays of a \mtau to a lighter-mass charged lepton and an $\omega$ vector meson 
is performed using \lumi of \epem annihilation data collected with  the \babar\ detector at the Stanford Linear 
Accelerator Center PEP-II storage ring. No signal is found, and the upper limits on the branching ratios are determined 
to be \UpperLimiteome and \UpperLimitmome at 90\% confidence level. 
\end{abstract}


\maketitle
In the Standard Model (SM) with massless neutrinos, lepton number 
is conserved separately for each generation. However the discovery of large neutrino 
mixing~\cite{NuOsc} requires that lepton flavor violation (LFV) occur, although 
decays involving charged LFV have not yet been observed. 
In minimal extensions of the SM that account for neutrino oscillations
by the seesaw mechanism of neutrino mass generation, the expected rates of LFV decays are  
too small to be observable. Thus the observation of neutrinoless decays 
like \tautolome, where $\ell = e$ or $\mu$, would be an unambiguous signature of new physics~\cite{Ilakovac:1999md},
while limits on this process provide constraints on theoretical models.

The search for \tautolome decays presented here uses data recorded by the \babar\ detector 
at the SLAC \pep2 asymmetric-energy \epem storage ring.
The data sample corresponds to an integrated luminosity \L = \lumion
recorded at an \epem center-of-mass (\CM) energy \roots = 10.58\gev, and 
\lumioff recorded at \roots = 10.54\gev. With an average cross section 
of $\sigma_{\eett}$ = (0.919$\pm$0.003) nb~\cite{kkxsec}, this corresponds to a 
sample of \ntaupair $\tau$-pair events.

The details of the \babar\ detector are described elsewhere~\cite{detector}.
Charged particles are reconstructed as tracks with a five-layer silicon vertex tracker (SVT) 
and a 40-layer drift chamber (DCH) inside a 1.5 T solenoidal magnet. 
An electromagnetic calorimeter (EMC) consisting of 6580 CsI(Tl) crystals 
is used to identify electrons and photons. A ring-imaging Cherenkov detector (DIRC) is used to identify 
charged pions and kaons. The flux return of the solenoid, instrumented with resistive plate chambers 
and limited streamer tubes, is used to identify muons.

The signature of the signal process is the presence of a $\ell\omega$ pair having an invariant mass
consistent with $m_\tau$ = 1.777\gevcc~\cite{bes} and a total energy equal to \roots/2 in the \CM frame,
along with other particles in the event whose properties are consistent with a \mtau decay.
Only the dominant decay mode of the $\omega$ meson (\omeppp) is used in this analysis.
The estimation of the background rate in the final sample comes from data only,
while samples of Monte Carlo (MC) simulated events are used to obtain the
signal reconstruction efficiency, the kinematic distributions of the signal and background events, to optimize the selection criteria
and to study systematic uncertainties in the signal efficiency. Control samples with two identified electrons 
in the event are used to study background contamination from radiative Bhabha scattering, since
the relevant cross-section is large, making it impractical to generate a sufficient number of simulated events.

The signal events are simulated with \kk~\cite{kk}, where one \mtau decays to $\ell^\pm \omega$ according to two body phase space
and the other \mtau decays according to measured branching fractions~\cite{Yao:2006px}
simulated with \tauola~\cite{tauola}. The $\mu^+\mu^-$ and \tautau background processes are generated using \kk and \tauola,
and \qqbar processes are generated using the \evtgen~\cite{Lange:2001uf} and \jetset~\cite{Sjostrand:1995iq} packages.
The detector response for the MC events is simulated using the \geant package~\cite{geant}.
Radiative corrections for signal and background processes are simulated using \photos~\cite{Golonka:2005pn}.

Events with four well-reconstructed tracks and zero total charge are selected, where no track pair is consistent with being 
a photon conversion in the detector material. Each event is divided into two hemispheres in the \CM frame 
by a plane perpendicular to the thrust axis~\cite{thrust}, calculated using all reconstructed charged and neutral particles.
The events having 3-1 topology, where the signal-side hemisphere contains three tracks and the tag-side contains 
one track, are selected. 

Photon candidates are required to have the measured energy in the EMC greater than 0.1\gev to 
reduce the background originating from the \epem colliding beams in the accelerator beam pipe.
Pairs of these photons are combined to form \piz candidates, with the invariant mass in the 
range $m(\gaga)$ $\in$ $[0.115, 0.150]$ \gevcc. The $\omega$ mesons are reconstructed from two oppositely charged pion candidates 
combined with a \piz, with the invariant mass in the range $m(\ppp)$ $\in$ $[0.760, 0.805]$ \gevcc.
In the \tautolome decay, two of the tracks in the signal-side hemisphere have the same charge.
Each of these two same-sign tracks is combined independently with the opposite-sign track and the neutral pion to form
two $\omega$ candidates. The candidate with invariant mass nearest to the nominal $\omega$ mass~\cite{Yao:2006px} is
considered to be the signal $\omega$.
The signal track that is not combined to form the $\omega$ candidate is required to have a momentum in the laboratory frame greater 
than 0.5\gevc and to be identified as an electron or muon as appropriate, using \babar\ particle identification 
techniques~\cite{Aubert:2002rg}.
The three tracks in the signal-side hemisphere are fitted to a common vertex, and the photons from the \piz are assumed to originate 
from this vertex. The reconstructed \piz candidate from the signal \mtau is constrained to the nominal \piz mass~\cite{Yao:2006px}.
The $\omega$ candidate is then combined with the lepton track to form the signal \mtau candidate.
The signal-side hemisphere may contain up to four photons
so as to allow hadronic split-offs from the pion tracks
in the EMC. Thus, there may be more than one \piz candidate,
resulting in multiple \mtau candidates.
In this case, the $\ell \omega$ combination with invariant mass closest to the nominal \mtau mass is accepted
as the signal \mtau candidate. From a sample of $1.6 \times 10^{6}$ generated signal MC events,
all the reconstructed signal candidates are verified to have correct association with the truth-matched signal \mtau decays.

Signal events are distinguished by two kinematic variables:
the beam-energy constrained mass (\mec) and the energy difference
$\DeltaE = E_{\ell} + E_{\omega} - \roots/2$, where $E_\ell$ and $E_{\omega}$ are energies of the lepton and
the $\omega$ in the \CM frame. The \mec is calculated from a fit to the reconstructed \mtau candidate
decay products with a constraint that the \mtau energy is equal to \roots/2 in the \CM frame.
These two variables are weakly correlated and have non-Gaussian tails due to initial and final state radiation.
For the signal MC events, the \mec distribution peaks at $m_\tau$,
while the \DeltaE distribution peaks close to but below zero, primarily due to photon energy
reconstruction effects producing a small negative offset in the reconstructed \mtau energy.
The peak positions ($\hat{m}_{\rm EC},~\Delta{\hat{E}}$) and standard deviations ($\sigma(\mec),~\sigma(\DeltaE)$) of 
the \mec and \DeltaE distributions for the reconstructed signal MC events 
are presented in Table~\ref{table1}. To study signal-like events, a large box (LB) is defined 
in the \mec vs. \DeltaE plane as: \mec $\in$ $[1.6, 2.0]$ \gevcc and \DeltaE $\in$ $[-0.8, 0.4]$ \gev. 
To avoid experimenter bias, the number and the properties of data events falling within 
the $\pm$3$\sigma$ rectangular region in the \mec-\DeltaE plane, defined as the signal box (SB), are neither 
used to optimize the selection criteria nor to study systematic effects.
The region inside the LB but outside the SB is called grand side band (GSB) and is used for estimation of the background 
contribution in the SB. The selection requirements are optimized to yield the lowest expected upper 
limit (UL)~\cite{Feldman:1997qc} derived from the events inside the SB under a background-only hypothesis.

To suppress non-\mtau backgrounds with radiation along the beam direction,
the polar angle of the missing momentum with respect to beam axis ($\theta_{\rm{miss}}$)
is required to lie within the detector acceptance: $\cos{\theta_{\rm{miss}}} \in [-0.76, 0.92]$. 
The total \CM momentum of all tracks and photon candidates in the tag-side must be less than 4.75\gevc.

The events are classified into four different categories depending on tag-side hemisphere properties: the particle 
identification for the track and the total neutral \CM energy in the hemisphere (\tagegam).
If the tag-side track is identified as an electron or a muon it is categorized as 
an $e-$tag or a $\mu-$tag. Otherwise it is categorized as an $h-$tag or a $\rho-$tag, depending on whether 
\tagegam is less than or greater than 0.2\gev. The $e-$tag events are not used in the final selection 
of \tautoeome candidates, but are used as the control sample to estimate the Bhabha contribution to this decay mode.

The tag-side hemisphere is expected to contain a SM \mtau decay characterized by the presence 
of one charged particle and one or two neutrinos. The missing mass due to the undetected neutrino(s) is reconstructed as
\Mnu = $(P^\mathrm{tag}_{\tau} - P^\mathrm{tag}_\mathrm{obs})^2$, where $P^\mathrm{tag}_{\tau}$ and 
$P^\mathrm{tag}_\mathrm{obs}$ are four-momenta in the \CM frame. The energy and momentum components of 
$P^\mathrm{tag}_{\tau}$ are (\roots/2, $\sqrt{(s/4 - {m_\tau}^2)} \cdot \hat{n}$) where $\hat{n}$ is the 
unit vector opposite in direction to the signal-side \mtau momentum and $P^\mathrm{tag}_\mathrm{obs}$ is the combined 
four-momentum in the \CM frame  of all the tracks and photon candidates observed in the tag-side hemisphere.
To reduce non-$\tau$ backgrounds, tag-dependent requirements on \Mnu are applied for the \tautomome candidates.
For $e-$tags and $\mu-$tags, \Mnu must be in the range $\in$ $[-2.0, 2.5] \gev^2/{c^4}$ whereas 
for $h-$tags and $\rho-$tags, \Mnu $\in$ $[-1.2, 2.0] \gev^2/{c^4}$ and \Mnu $\in$ $[-2.0, 0.5] \gev^2/{c^4}$, respectively.
For the \tautomome candidates, the ratio \logmisspt in the \CM frame is required to be greater 
than 0.061, where \ptmiss is the component of the missing momentum of the event transverse to the beam direction.
For \tautoeome candidates, \logmisspt is required to be greater than 0.034.

\begin{figure*}[htbp]
\resizebox{0.96\textwidth}{0.28\textheight}{
\includegraphics{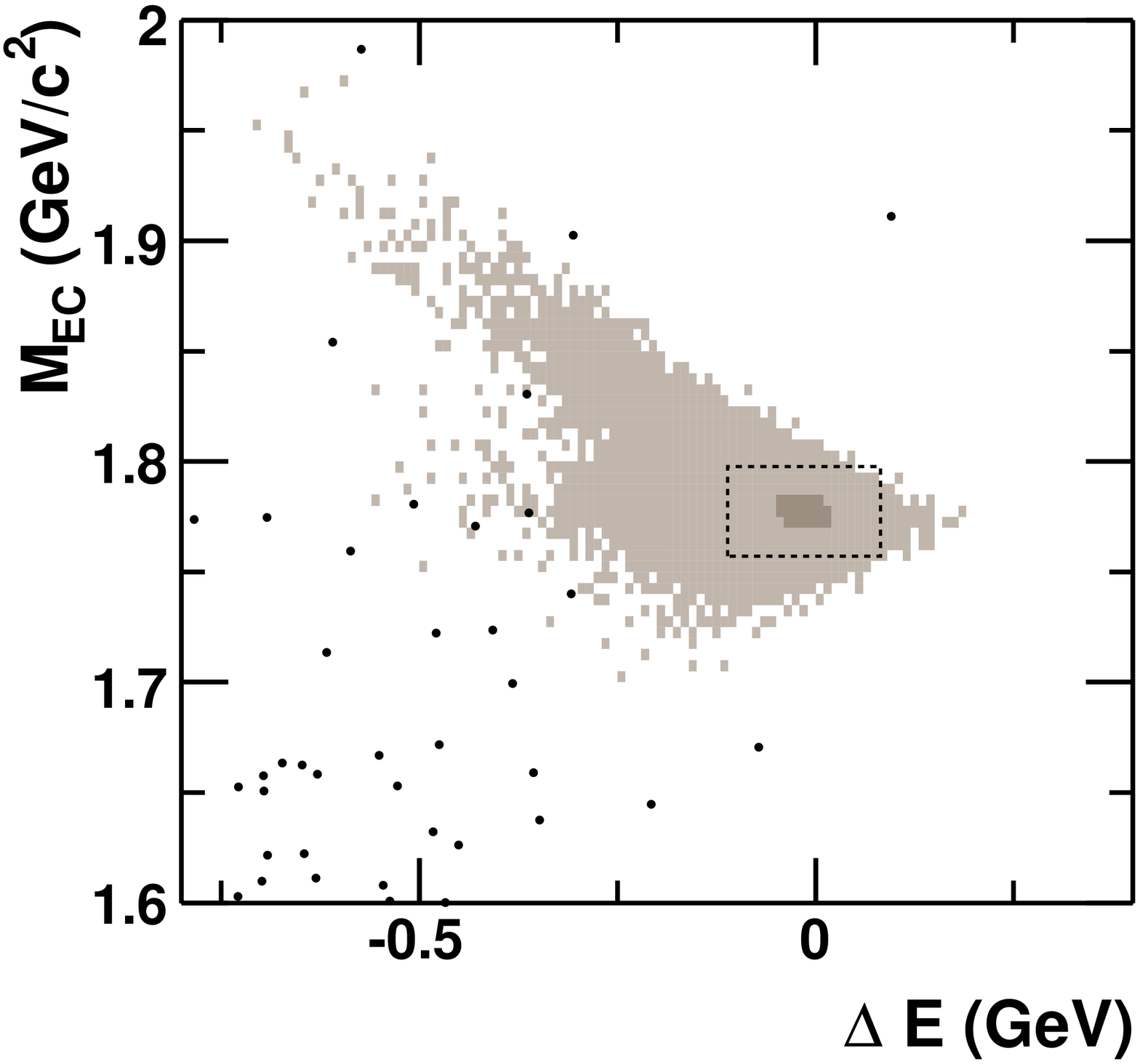} 
\includegraphics{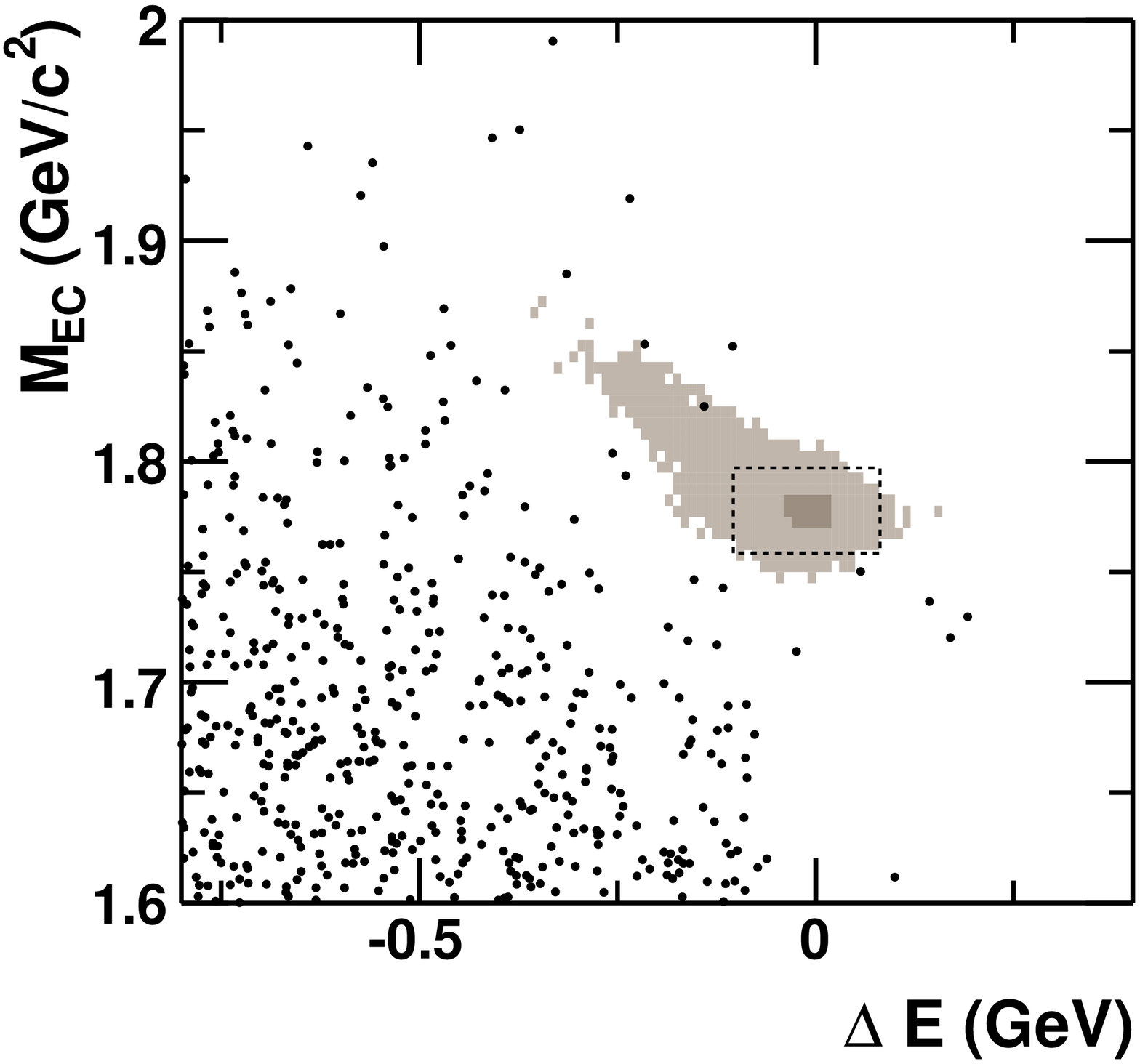}
}
\caption{The selected candidates (dots) inside the large box region of the \mec-\DeltaE plane for \tautoeome (left plot) 
           and \tautomome (right plot) decays. The $\pm3\sigma$ signal box is shown by a dashed rectangle. The dark and light
           shading indicates contours containing 50\% and 90\% of the selected MC signal events, respectively. The signal box
           contains 67\% of the selected MC signal events for \tautoeome and 77\% for \tautomome decay.} 

\label{fig1}
\end{figure*}

After applying all the selection criteria for \tautoeome and \tautomome decays, the number of data events surviving inside 
the GSB  are 39 and 502, respectively, as shown in Fig.~\ref{fig1}. 
The number of background events in the MC and control samples, in the same region and passing the same set of 
requirements as data, is 35$\pm$6 for \tautoeome and 564$\pm$26 for \tautomome decay.
Out of these MC background events in the \tautoeome decay, the dominant contributions are from $\qqbar$ (54\%) and \tautau (34\%);
the rest arise from radiative Bhabha scattering. About 92\% of the background in \tautomome decay is 
from \tautau events; within this category, 94\% are due to the decay \tautoomega, where one of the charged pions is 
mis-identified as a muon. The number of background events in the \tautomome sample is more than \tautoeome because 
of the larger mis-identification rate for a pion track to be identified as a muon than an electron. 

The number of expected background events in the SB is extracted from
an unbinned maximum likelihood fit to the \mec and \DeltaE distributions of data events
inside the GSB, using a two-dimensional probability density function (PDF) made of a linear 
combination of PDFs representing each background component, \epem, \mumu, \tautau and \qqbar. 
The MC event samples are used to determine each component PDF but the one
describing radiative Bhabha events, for which a data control sample is
used. Each PDF is obtained by interpolating the two-dimensional
binned distribution of its respective sample using Gaussian
weight terms that are fit with an adaptive kernel estimation procedure~\cite{keys}.
The expected background normalization is fixed to the amount of data events in the GSB, while 
the relative yields of the different background components are fitted to the background shape.
The numbers of background events expected from this fit for various regions around the SB are compared 
with the numbers observed in Table~\ref{table2}, and they confirm that the backgrounds in the data are adequately modeled.
\renewcommand{\multirowsetup}{\centering}
\newlength{\LL}\settowidth{\LL}{$8047$}
\begin{table*}[htbp]
\begin{center}
\caption{The peak positions and standard deviations of the \mec and \DeltaE distributions, obtained from the fit to the signal MC events.
Also shown are the reconstruction efficiencies (\eff), the number of expected background (exp.) events and the observed (obs.) events 
inside the signal box, and the resulting upper limits at 90\% confidence level (C.L.) including the systematic uncertainties.}
\label{table1}
\begin{ruledtabular}
\renewcommand{\arraystretch}{1.10}
{\scalebox{0.99}{
\begin{tabular}{lccccccccc}
Decay modes  & $\hat{m}_{\rm EC}$  & $\sigma(\mec)$& $\Delta{\hat{E}}$ & $\sigma(\DeltaE)$ & \eff &\multicolumn{2}{c}{SB events}&\multicolumn{2}{c}{UL (\tenseven)}\\\cline{2-10}
             &   \mevcc  & \mevcc       & \mev    &\mev           & (\%)   &exp.&obs.                             &exp.&obs.  \\\hline
\tautoeome   &1777.4$\pm$0.1 & 6.8$\pm$0.1 & -14.4$\pm$0.3 & 32.2$\pm$0.3 & 2.96$\pm$0.13 & 0.35$\pm$0.06 & 0 & 1.4 & 1.1   \\\hline 
\tautomome   &1777.7$\pm$0.1 & 6.4$\pm$0.1 & -11.2$\pm$0.2 & 30.9$\pm$0.3 & 2.56$\pm$0.16 & 0.73$\pm$0.03 & 0 & 1.7 & 1.0   \\
\end{tabular}
}}
\end{ruledtabular}
\end{center}
\end{table*}
\renewcommand{\multirowsetup}{\centering}
\begin{table*}[!htbp]
\begin{center}
\caption{
The expected number of background events obtained from the fit to \mec${-}$\DeltaE distributions  
within the $\pm(3{-}5)\sigma$, $\pm(5{-}7)\sigma$, $\pm(7{-}9)\sigma$, $\pm(9{-}11)\sigma$ and the 
combined $\pm(3{-}11)\sigma$ nested rectangular regions centered around the signal box. Also shown are
the number of observed events inside the corresponding regions.}
\label{table2}
\begin{ruledtabular}
\renewcommand{\arraystretch}{1.10}
{\scalebox{0.99}{
\begin{tabular}{lcccccc}
Decay modes  & \# of events       & $\pm(3{-}5)\sigma$  & $\pm(5{-}7)\sigma$   & $\pm(7{-}9)\sigma$   & $\pm(9{-}11)\sigma$   &  $\pm(3{-}11)\sigma$ \\ \hline
\tautoeome   & expected           & 0.6$\pm$0.1  & 1.0$\pm$0.2   &  1.4$\pm$0.2  &  1.9$\pm$0.3   & 4.9$\pm$0.8        \\ \cline{2-7} 
             & observed           &   0          &    0          &   1           &    2           & 3                \\ \hline
\tautomome   & expected           &1.9$\pm$0.1   & 3.9$\pm$0.2   & 6.7$\pm$0.3   & 12.1$\pm$0.5   & 24.6$\pm$1.1        \\ \cline{2-7}
             & observed           & 2            &    3          &     7         & 10             & 22                   \\
\end{tabular}
}}
\end{ruledtabular}
\end{center}
\end{table*}

Systematic uncertainties on the signal efficiency and the estimated
background are considered in this measurement.
The uncertainty due to knowledge of the efficiencies for 
trigger, for the tracking, and in the beam energy scale and spread is 1.4\% for both decay modes.  
An uncertainty of 2.0\% originates from uncertainties on the lepton
track momentum and on the photon energy scale and resolution, which
affect the position and spread of the \DeltaE and \mec distributions.
There is a 3.3\% uncertainty in the
\piz reconstruction efficiency, the uncertainty in lepton identification is
1.1\% for electrons and 4.5\% for muons, and there is a 1\% uncertainty on 
the number of \mtau pairs produced. After combining these individual contributions in quadrature, the total systematic 
uncertainty on efficiency is 4.4\% for \tautoeome and 6.2\% for \tautomome. 
The uncertainties on background estimation are determined by the background fit errors.
The uncertainty due to MC statistics is negligible. 

The signal is simulated according to the two body phase space, i.e. with a 
uniform distribution of the cosine of the helicity angle with respect to the \mtau spin. 
Since \mtau pairs are produced with spin correlation, the event selection efficiency may
be sensitive to the helicity angle distribution of the \tautolome
decay, which depends on the model of the LFV interaction~\cite{flatphasespace}. This effect
is simulated by weighting the generated events to match the
helicity angle distributions of both $V-A$ and $V+A$ interactions and its
consequences on the measured upper limit is found to be
negligible.

The upper limits for \tautolome decays are calculated using 
$\BR^{90}_{UL}=N^{90}_{UL}/(2\L\sigma_{\tau\tau}\BR\eff)$,
where $N^{90}_{UL}$ is the 90\% C.L. upper limit 
on the number of signal events inside the SB, \BR\ is the branching fraction~\cite{Yao:2006px} 
of the decay \omeppp (\ptogg) and \eff is the reconstruction efficiency of the signal decay mode under consideration.
The expected and observed upper limits, including all contributions from systematic 
uncertainties, are calculated using the technique of Cousins and Highland~\cite{Cousins:1992qz}
with the implementation of Barlow~\cite{Barlow:2002bk}. 
No signal is found, and the upper limits on the branching ratios are determined 
to be \UpperLimiteome and \UpperLimitmome at 90\% confidence level, as shown in  Table~\ref{table1}. 

We are grateful for the excellent luminosity and machine conditions
provided by our \pep2\ colleagues, 
and for the substantial dedicated effort from
the computing organizations that support \babar.
The collaborating institutions wish to thank 
SLAC for its support and kind hospitality. 
This work is supported by
DOE
and NSF (USA),
NSERC (Canada),
CEA and
CNRS-IN2P3
(France),
BMBF and DFG
(Germany),
INFN (Italy),
FOM (The Netherlands),
NFR (Norway),
MIST (Russia),
MEC (Spain), and
STFC (United Kingdom). 
Individuals have received support from the
Marie Curie EIF (European Union) and
the A.~P.~Sloan Foundation.


\end{document}